\documentclass[aps, prl,twocolumn,preprintnumbers,amsmath,amssymb]{revtex4}
\usepackage{graphicx}
\begin{document}
\title{Notes on the Vollhardt "invariant" and phase transition in the helical itinerant magnet MnSi}

\author{S.M. Stishov}
\email{sergei@hppi.troitsk.ru}
\author{A.E. Petrova}
\affiliation{Institute for High Pressure Physics of RAS, Troitsk, Moscow, Russia}

\begin{abstract}
In this paper we argue that rounded "hills" or "valleys" demonstrated by the heat capacity, thermal expansion coefficient, and elastic module are indications of a smeared second order phase transition, which is flattened and spread out by the application of a magnetic field. As a result, some of the curves which display a temperature dependence of the  corresponding quantities cross almost at a single point. Thus, the Vollhardt crossing point should not be identified with any specific energy scale. The smeared phase transition in MnSi preceding the helical first order transition most probably corresponds to the planar ferromagnetic ordering, with a small or negligible correlation between planes. At lower temperatures, the system of ferromagnetic planes becomes correlated, acquiring a helical twist.   
\end{abstract}
\maketitle

\section{Introduction}
It  has been known that at the magnetic phase transition in MnSi, some  thermodynamic quantities  behave in an unusual way. The heat capacity, thermal expansion,  and elastic moduli form prominent rounded "hills" or "valleys" with first order transition features on their low temperature sides~\cite{1,2,3,4,5,6}. An interpretation of this "topography" as an indication of helical fluctuations does not answer all the questions~\cite{7,8,9}. In particular, why do these fluctuations start to decay on approaching the phase transition?  
On the other hand, it is tempting to suppose that these features can be evidence of some special kind of phase transformation. Indeed, as was pointed out in Ref.~\cite{6}: " \ldots the magnetic phase transition in MnSi, occurring at 28.8 K, is just a minor feature of the global transformation that is marked by the rounded maxima or minima of heat capacity, thermal expansion coefficient, sound velocities and absorption, and the temperature derivative of resistivity".

\begin{figure}[htb]
\includegraphics[width=80mm]{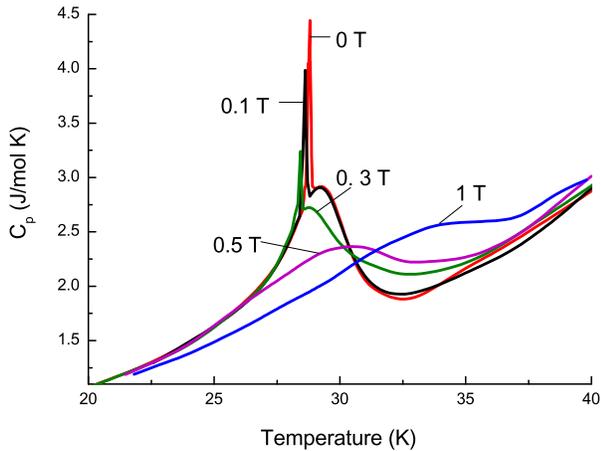}
\caption{\label{fig1} (Color online) Heat capacity of MnSi as a function of temperature and magnetic field (drawn after data of Ref.~\cite{5}).}
\end{figure}

\begin{figure}[htb]
\includegraphics[width=80mm]{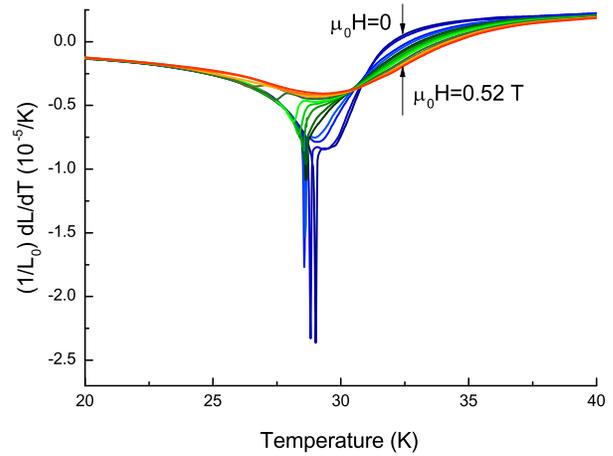}
\caption{\label{fig2} (Color online) Thermal expansion coefficient of MnSi as a function of temperature and magnetic field  (drawn after data of Ref.~\cite{12}).}
\end{figure}

\section{Vollhardt Crossing points}
\begin{figure}[htb]
\includegraphics[width=80mm]{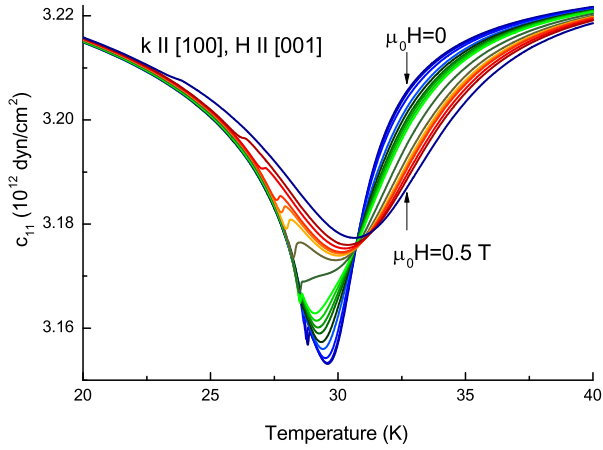}
\caption{\label{fig3} (Color online) Elastic modulus $c_{11}$ of MnSi as a function of temperature and magnetic field (drawn after data of Ref.~\cite{13}).}
\end{figure}

Recently the attention of some researchers in the field has been turned   
to the Vollhardt observations that in many correlated systems the specific heat curves when plotted for different values of some thermodynamic variable, e.g. magnetic field, cross at some temperatures ~\cite{10}. The heat capacity crossing point in MnSi (see Fig.~\ref{fig1}), named a Vollhardt invariant in Ref.~\cite{11}, supposedly identifies an intrinsic energy scale of the system ~\cite{11} and is considered an important piece of verification for the Brazovskii scenario of phase transition in MnSi ~\cite{9}.  In this connection, it should be pointed out that results obtained in Ref.~\cite{12} question the applicability of a model of a fluctuation - induced first order phase transition for MnSi.  
	In fact, the crossing points are also observed in the thermal expansion and elastic behavior of MnSi (see Figs.~\ref{fig2} and ~\ref{fig3}). 

From Figs.~\ref{fig1} - ~\ref{fig3} one can get the  relationships:
\begin{equation}\label{eq1}
\left( \dfrac{\partial C_{p}}{\partial H•}\right) _{T^{*}}=T\left(\dfrac{\partial ^{2}M}{\partial T^{2}} \right) _{T^{*}}\approx TH\left( \dfrac{\partial^{2} \chi}{\partial T^{2}}\right) _{T^{*}}=0
\end{equation}
\begin{equation}\label{eq2}
\left( \dfrac{\partial K}{\partial H•}\right) _{T^{*}}=T\left(\dfrac{\partial ^{2}M}{\partial V^{2}} \right) _{T^{*}}\approx HV\left( \dfrac{\partial^{2}\chi}{\partial V^{2}}\right) _{T^{*}}=0
\end{equation}
\begin{equation}\label{eq3}
\left( \dfrac{\partial \alpha}{\partial H•}\right) _{T^{*}}=\dfrac{1}{V}\left(\dfrac{\partial ^{2}M}{\partial p \partial T} \right) _{T^{*}}\approx \dfrac{H}{V}\left( \dfrac{\partial^{2}\chi}{\partial p \partial T}\right) _{T^{*}}=0
\end{equation}

\begin{figure}[htb]
\includegraphics[width=80mm]{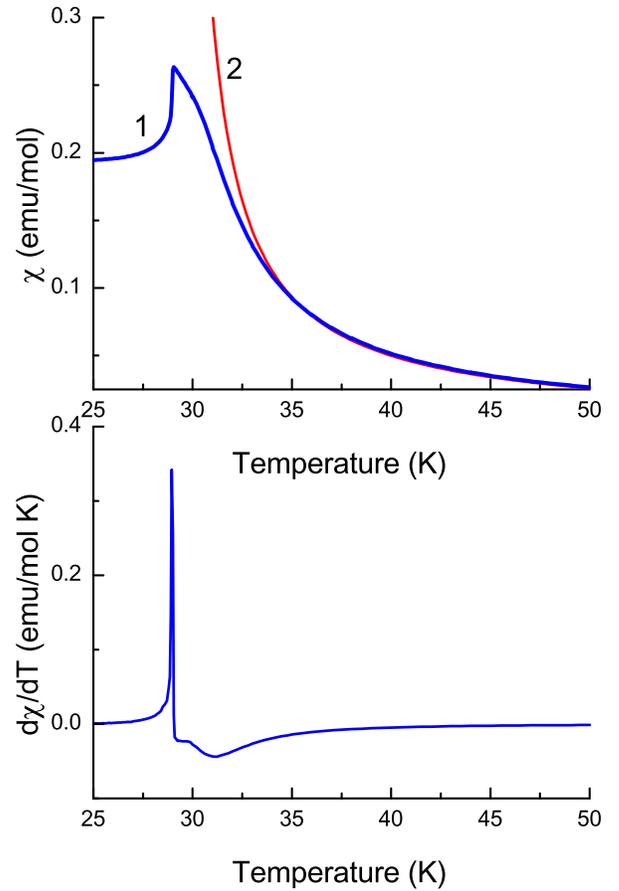}
\caption{\label{fig4} (Color online) Upper plot. Magnetic susceptibility $\chi=M/H$ of MnSi
as a function of temperature (1). Extrapolation of high temperature data according to the Curie-Weiss law (2).
Low plot. Temperature derivative of magnetic susceptibility $d\chi/dT$ as a function of temperature. Zero value of $d^2\chi/dT^2$ is clearly seen at $\sim 31$ K (drawn after data of Ref.~\cite{4}).}
\end{figure}

Here $C_{p}$ - heat capacity, $M$ - magnetization, $\chi=M/H$  – magnetic susceptibility, $K=-V(\partial P / \partial V)_{T}$ - bulk modulus, 
$\alpha=(1/V) (\partial  V /  \partial T)_{P}$ - thermal expansion coefficient.
The equations~\ref{eq1}--\ref{eq3} show that the magnetic susceptibility curve experiences an inflection at some particular temperature $T^{*}$ situated slightly above $T_{c}$ independent of the way of approaching this temperature.  Note that the behavior of magnetic susceptibility in the vicinity of the phase transition point in MnSi in the paramagnetic phase is highly unusual (Fig.~\ref{fig4}). Indeed, one would expect magnetic susceptibility to diverge  at $T_{c}$ according to the Curie-Weiss law, with a power exponent $\gamma=1$ or higher in case of significant fluctuation contributions, but it did not happen. The magnetic susceptibility curve of MnSi inflects at $T_{*}$ and reaches a finite value at $T_{c}$. 
   
\begin{figure}[htb]
\includegraphics[width=80mm]{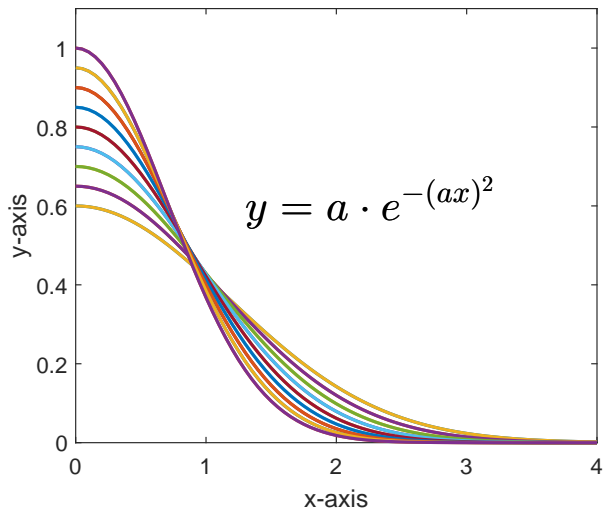}
\caption{\label{fig5} (Color online) Function $y=a\cdot exp(-a^2 x^2)$ at different values of width $a$.}
\end{figure}

 Coming back to the Vollhardt observation, one may conclude that the crossing phenomena could occur if the corresponding “hills” and ”valleys” would flatten and spread out under influence of a field, which is magnetic in this case.  This situation is modeled in Fig.~\ref{fig5}, where the function $ y=a\cdot exp(-a^2 x^2)$ is depicted at different values of the width $a$. Note that
$\int_{0}^{\infty}a\cdot exp(-a^2 x^2)dx = \dfrac{\pi^{1/2}}{2}$, i.e. the integral value does not depend on the width of the function.

As is seen in Fig.~\ref{fig5} the curves with $a=1, 0.95, 0.9, 0.85, 0.8, 0.75$, 
0.7 practically cross at a single point. Then at $a < 0.7$ the curves started to deviate from this trend. This clearly resembles what one can see in Figs.~\ref{fig1} - ~\ref{fig3}.  Based on this analogy, one may suggest that a magnetic field smears out the "hill" and "valley" anomalies in MnSi, not strongly affecting their integral values.

\begin{figure}[htb]
\includegraphics[width=80mm]{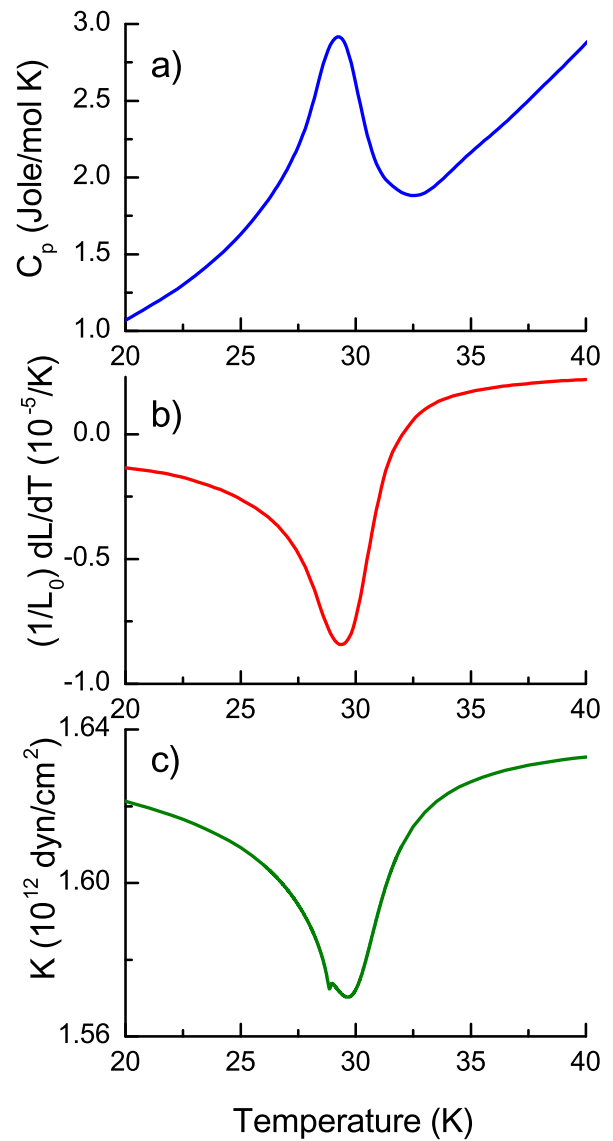}
\caption{\label{fig6} (Color online) Heat capacity a) and linear thermal expansion b) of MnSi as functions of temperature at H=0.  First order helical transition peaks are removed (cf. Figs.~\ref{fig1} and ~\ref{fig2}). Bulk modulus  c) of MnSi as a function of temperature at H=0. A small first order spike is seen on the left of the minimum (drawn after data of Ref.~\cite{4,12,13}). Note that the volume and entropy change at the first order phase transition in MnSi are extremely small ($\Delta V/V\approx10^{-6}$, $\Delta S/R\approx 10^{-4}$)~\cite{4,5} and can not influence much the anomalies shown in this figure.} 
\end{figure}

\section{Discussion}
In further discussion we intentionally ignored various theoretical and heuristic constructions devoted to various aspects of the physics of MnSi, trying to  base exceptionally on the experimental thermodynamic data, characterizing the magnetic phase transition in MnSi.
Then, extending the supposition that the maxima and minima in $C_{p}$, $\alpha=(1/L)(dL/dT)$ and $c_{11}$ correspond to some kind of a phase transition, which is smeared out by the magnetic field, one may think of the ferromagnetic order as a conjugate parameter to this field. Indeed, a continuous phase transition is smoothed by an application of a conjugate field ~\cite{14}, although the decaying thermodynamic anomalies can be still seen until they are completely wiped out by a strong field. 
Therefore, the implication is that the mentioned above maxima and minima manifest a so-called smeared second order phase transition (see, for instance ~\cite{14,15}). In the XY spin subsystem in MnSi this transition obviously corresponds to planar ferromagnetic ordering, with a small or negligible correlation between planes. At lower temperature the system of ferromagnetic planes becomes correlated, acquiring a helical twist.  This situation is beautifully illustrated by Fig.~\ref{fig6}, where heat capacity and linear thermal expansion of MnSi are depicted with the first order peaks removed for better viewing.
The most instructive is the behavior of the bulk modulus of MnSi (Fig.~\ref{fig6}c), which only shows a small first order jump on the background of the large anomaly, demonstrating a tendency to volume instability ~\cite{6}.  The smearing possibly arises from a generic inhomogeneity of MnSi. The latter could be a local deviation from the stoichiometry facilitated by the polyvalent character of Mn.   

\section{Conclusion}
We argue that rounded "hills" or "valleys" demonstrated by the heat capacity, thermal expansion and elastic module are indications of a smeared second order phase transition, which is flattened and spread out by the application of a magnetic field. As a result, some of the curves displaying a temperature dependence of the corresponding quantities may cross almost at a single point, as illustrated in Fig.~\ref{fig5}. Thus, the Vollhardt crossing point should not be identified with any specific energy scale as it was suggested in Ref.~\cite{9,11}. The smeared phase transition in MnSi preceding the helical first order transition most probably corresponds to the planar ferromagnetic ordering with a small or negligible correlation between planes. At lower temperatures, the system of ferromagnetic planes becomes correlated, acquiring a helical twist.   
\section{Acknowledgements}
SMS (data analysis, writing the paper) and AEP (data analysis) greatly appreciate financial support   of  the Russian Foundation for Basic Research (grant 15-02-02040), Program of the Physics Department of RAS on Strongly Correlated Electron Systems and Program of the Presidium of RAS on Strongly Compressed Matter. AEP is also grateful to the Russian Science Foundation (14-22-00093) for financial support. SMS and AEP thank Damon and Ellie Giovanielli for valuable remarks.
	
\end{document}